\documentclass[nofootinbib,reprint,superscriptaddress,pdftex]{revtex4-1}
\usepackage{amsmath,amssymb}
\usepackage[dvips]{graphicx,color}
\usepackage[pdftex,colorlinks]{hyperref}
\numberwithin{equation}{section}
\begin{document}

\title{Quantum Brownian motion of multipartite systems and their entanglement dynamics}
\author{C.~H.~Fleming}
\affiliation{Joint Quantum Institute and Department of Physics, University of Maryland, College Park, Maryland 20742}
\author{Albert Roura}
\affiliation{Max-Planck-Institut f\"ur Gravitationsphysik (Albert-Einstein-Institut),
Am M\"uhlenberg 1, 14476 Golm, Germany}
\author{B.~L.~Hu}
\affiliation{Joint Quantum Institute and Department of Physics, University of Maryland, College Park, Maryland 20742}

\date{\today}

\begin{abstract}
We solve the model of $N$ quantum Brownian oscillators linearly coupled to an environment of quantum oscillators at finite temperature,
with no extra assumptions about the structure of the system-environment coupling.
Using a compact phase-space formalism, we give a rather quick and direct derivation of the master equation and its solutions for general spectral functions and arbitrary temperatures.
Since our framework is intrinsically nonperturbative, we are able to analyze the entanglement dynamics of two oscillators coupled to a common scalar field in previously unexplored regimes, such as off resonance and strong coupling.
\end{abstract}

\maketitle

\tableofcontents

\section{Introduction}
\subsection{NQBM}
In Ref.~\cite{QBM} exact solutions were obtained for the problem of one Brownian oscillator linearly coupled to an environment of quantum oscillators, or quantum Brownian motion (QBM).
It was briefly mentioned how the phase-space formalism therein rather trivially generalized to the problem of $N$ Brownian oscillators with arbitrary spectral-density function, or {NQBM}.
In this work we make explicit this generalization and clarify details which become more important in multipartite systems.

An important discovery made in Ref.~\cite{QBM} was that previous derivations of the QBM master equation were invalid for nonlocal dissipation,
due to a subtle issue concerning boundary conditions of integro-differential equations.
This issue remains when considering multiple system oscillators and, thus, applies also to NQBM treatments prior to Ref.~\cite{QBM}.
Another important result there was the calculation of exact, analytic results for sub-Ohmic, Ohmic, and supra-Ohmic couplings.
The suitability of our QBM formalism for such exact calculations remains in the NQBM case. However, as a nontrivial application of our formalism we study here
the case of several detectors coupled to a relativistic-field environment, where the spectral density is far more complicated than a simple power law.

NQBM was previously considered in Ref.~\cite{Chou08},
but only for couplings such that the environment is effectively coupled only to one system degree of freedom.
Ref.~\cite{Lin09} went into great detail for resonant system oscillators, but otherwise treated the problem generally enough to model local oscillators interacting via a relativistic field.
The most general regime was briefly discussed in Ref.~\cite{QBM} and has since been covered by Ref.~\cite{Anastopoulos10} in some detail.
In this work we develop the formalism and expressions which allowed us to obtain the exact analytic solutions in Ref.~\cite{QBM}.

\subsection{Entanglement Dynamics}
In the second part of the paper we apply our formalism to investigate the {entanglement dynamics} of local oscillators in a common scalar-field environment.
The scalar-field environment provides a source of noise which is correlated, not only in time, but also in space.
This detail has a large effect upon the entanglement dynamics of the oscillators.
The regime we consider is very general and encapsulates those considered in other treatments.
The study in Ref.~\cite{Paz08} essentially corresponds to the case of two local and resonant oscillators at precisely the same location within their environment,
such that their environmental influences were identical.
Ref.~\cite{Lin09} 
carried out a perturbative analysis of 
the entanglement dynamics of local and resonant oscillators at separate locations in a scalar-field environment,
so that their environmental influences were intricately correlated according to their separation.
On the other hand, Ref.~\cite{Anastopoulos10} considered non-resonant oscillators but with a generalized Ohmic coupling rather than a field environment exhibiting nontrivial spatial correlations.

\subsection{Organization}
In Sec.~\ref{sec:NQBM} we provide a fairly self-contained derivation and analysis of the $N$-oscillator solutions.
This includes solutions to the Langevin equations and master equation.
Due to the matrix formulation in Ref.~\cite{QBM}, most of the useful relations therein can be generalized quite naturally.
In Sec.~\ref{sec:refine} we develop relations for the system evolution and late-time covariance which are especially useful in multipartite systems.
Finally in Sec.~\ref{sec:Detectors} we apply this formalism to the problem of a pair of local oscillators residing in the same scalar field,
but with some finite separation between them.

Throughout the paper we use natural units with $\hbar=c=1$.

\section{N-QBM}\label{sec:NQBM}
\subsection{The Lagrangian}

Our model is that of a continuous and linear system with finite and countable degrees of freedom, with Lagrangian $L_\mathrm{sys}(\mathbf{X},\dot{\mathbf{X}})$,
bilinearly coupled, via a Lagrangian $L_\mathrm{int}(\mathbf{X},\mathbf{x})$,
to a linear environment with an infinite (and possibly continuous) number of degrees of freedom, with Lagrangian $L_\mathrm{env}(\mathbf{x},\dot{\mathbf{x}})$.
\begin{align}
L =&\; L_\mathrm{sys}(\mathbf{X},\dot{\mathbf{X}}) + L_\mathrm{env}(\mathbf{x},\dot{\mathbf{x}}) + L_\mathrm{int}(\mathbf{X},\mathbf{x}) + L_\mathrm{ren}(\mathbf{X}) \, , \\
L =&\; \frac{1}{2} \left( \dot{\mathbf{X}}^\mathrm{T} \mathbf{M} \, \dot{\mathbf{X}} - \mathbf{X}^\mathrm{T} \mathbf{C} \, \mathbf{X} \right)
+ \frac{1}{2} \left( \dot{\mathbf{x}}^\mathrm{T} \mathbf{m} \, \dot{\mathbf{x}} - \mathbf{x}^{\!\mathrm{T}} \mathbf{c} \, \mathbf{x} \right) \nonumber \\
& -\mathbf{x}^{\!\mathrm{T}} \mathbf{g} \, \mathbf{X} + L_\mathrm{ren}(\mathbf{X}) \, . 
\end{align}
We assume that the spring constant matrices $\mathbf{C}, \mathbf{c}$ as well as the mass matrices $\mathbf{M}, \mathbf{m}$ are real and positive definite, and can be considered in general to be symmetric.
If necessary, one can relax the positivity condition and even consider time-dependent mass matrices, spring constant matrices and system environment coupling matrix $\mathbf{g}$ \cite{Fleming11T}.
To ensure that the free and interacting system are similar in behavior, we will also include the renormalization $L_\mathrm{ren}(\mathbf{X})$.
Our choice of renormalization will be equivalent to inserting the entire system-environment interaction in the square of the potential:
\begin{align}
L =&\; \frac{1}{2} \left( \dot{\mathbf{X}}^\mathrm{T} \mathbf{M} \, \dot{\mathbf{X}} - \mathbf{X}^\mathrm{T} \mathbf{C} \, \mathbf{X} \right) \\
& + \frac{1}{2} \left( \dot{\mathbf{x}}^\mathrm{T} \mathbf{m} \, \dot{\mathbf{x}} - \left[ \mathbf{x} + \mathbf{c}^{-1} \mathbf{g} \, \mathbf{X} \right]^\mathrm{T} \!\! \mathbf{c} \left[ \mathbf{x} + \mathbf{c}^{-1} \mathbf{g} \, \mathbf{X} \right] \right) \, , \nonumber
\end{align}
since this keeps the phenomenological system-system couplings from changing.
Ref.~\cite{Lin09} chose to only renormalize the environmentally-induced self interactions or diagonal terms, which are the most cutoff sensitive terms.
In some ways that choice is more analogous to quantum electrodynamics.

\subsection{The Langevin Equation}

For the linear system there are several formalisms which produce the same Langevin Equation.
The most direct is via integrating out environment degrees of freedom in the Heisenberg equations of motion \cite{FordOconnell88} and then considering the symmetrized moments.
Another is to consider the characteristic curves of the system + environment's Fokker-Plank equation \cite{Fleming11T}.
Finally, one can integrate out both the environment degrees of freedom and the relative system coordinate $\boldsymbol{\Delta} = \mathbf{X} - \mathbf{X}'$,
while leaving only the average system coordinate $\boldsymbol{\Sigma} = (\mathbf{X} - \mathbf{X}')/2$,
in the double path integral of the reduced system propagator in the influence functional formalism \cite{CRV03}.
In general (for nonlinear systems) there is no necessary correspondence between these formalisms and only the first may be well defined, but here the Langevin equation is simply
\begin{align}
& \mathbf{M} \ddot{\mathbf{X}}(t) + (\mathbf{C} +\boldsymbol{\delta} \mathbf{C}) \mathbf{X}(t) + 2\int_0^t \!\! d\tau \, \boldsymbol{\mu}(t,\tau) \, \mathbf{X}(\tau) = \boldsymbol{\xi}(t) \, ,
\end{align}
where $\boldsymbol{\delta} \mathbf{C}$ is a renormalization of the system interaction, $\boldsymbol{\mu}(t,\tau)$ is the \emph{dissipation kernel},
and $\boldsymbol{\xi}(t)$ a Gaussian stochastic force with vanishing mean and correlation
\begin{eqnarray}
\left\langle \boldsymbol{\xi}(t) \, \boldsymbol{\xi}^\mathrm{T}(\tau) \right\rangle_{\boldsymbol{\xi}} &=& \boldsymbol{\nu}(t,\tau) \, ,
\end{eqnarray}
characterized by the real \emph{noise kernel} $\boldsymbol{\nu}(t,\tau)$.
In the phase-space representation, $\boldsymbol{\xi}(t)$ can be formally interpreted as a Gaussian stochastic process and $\mathbf{X}(t)$ as an unraveling of the characteristics which evolve the quantum state.

In general, for any Gaussian environment (and not only the one we model here) the two kernels are constrained by the fact that the environment correlator
\begin{eqnarray}
\boldsymbol{\alpha}(t,\tau) &=& \boldsymbol{\nu}(t,\tau) + \imath \, \boldsymbol{\mu}(t,\tau) \, ,
\end{eqnarray}
must be Hermitian and positive definite \cite{QOS}.
Specifically, for a constant bilinear coupling to a thermal reservoir of harmonic oscillators we have stationary kernels
\begin{align}
\boldsymbol{\mu}(t,\tau) &= - \mathbf{g}^\mathrm{T} \mathbf{m}^{\!-\frac{1}{2}} \frac{ \sin\!\left(\boldsymbol{\omega}[t\!-\!\tau]\right) }{ 2 \, \boldsymbol{\omega} } \mathbf{m}^{\!-\frac{1}{2}} \mathbf{g} \, , \\
\boldsymbol{\nu}(t,\tau) &= + \mathbf{g}^\mathrm{T} \mathbf{m}^{\!-\frac{1}{2}} \frac{ \coth\!\left( \frac{\boldsymbol{\omega}}{2T} \right) \cos\!\left(\boldsymbol{\omega}[t\!-\!\tau]\right) }{2 \, \boldsymbol{\omega}} \mathbf{m}^{\!-\frac{1}{2}} \mathbf{g} \, , \\
\boldsymbol{\omega}^2 & \equiv \mathbf{m}^{\!-\frac{1}{2}} \mathbf{c} \, \mathbf{m}^{\!-\frac{1}{2}} \, , 
\end{align}
with temperature-independent dissipation due to linearity in the field coupling.

For the positive-temperature environment, dissipation may also be represented via the positive-definite \emph{damping kernel}
defined by the relation $\boldsymbol{\mu}(t,\tau) = - \frac{\partial}{\partial t} \boldsymbol{\gamma}(t,\tau)$, such that
\begin{equation}
\boldsymbol{\gamma}(t,\tau) = +\mathbf{g}^\mathrm{T} \mathbf{m}^{\!-\frac{1}{2}} \frac{ \cos\!\left(\boldsymbol{\omega}[t\!-\!\tau]\right) }{ 2 \, \boldsymbol{\omega}^2 } \mathbf{m}^{\!-\frac{1}{2}} \mathbf{g} \, ,
\end{equation}
and with which the Langevin equation can be expressed as
\begin{align}
\mathbf{M} \ddot{\mathbf{X}}(t) + 2 \! \int_0^t \!\! d\tau \, \boldsymbol{\gamma}(t,\tau) \, \dot{\mathbf{X}}(\tau) + \mathbf{C} \mathbf{X}(t)  & \label{eq:langevin2} \\
+ 2 \, \boldsymbol{\gamma}(t) \, \mathbf{X}_0 &= \boldsymbol{\xi}(t)\, , \nonumber
\end{align}

The noise and damping kernels satisfy then the fluctuation-dissipation relation (here in the Fourier domain)
\begin{align}
\tilde{\boldsymbol{\nu}}(\omega) &= \tilde{\kappa}(\omega)  \, \tilde{\boldsymbol{\gamma}}(\omega) \, , \label{eq:FDR} \\
\tilde{\kappa}(\omega) & \equiv \hbar \omega \, \coth\!\left( \frac{\hbar \omega}{2 k_\mathrm{B} T} \right) \, , \label{eq:kappa}
\end{align}
with the Fourier transform defined
\begin{equation}
\tilde{f}(\omega) \equiv \int_{\!-\infty}^{+\infty} \!\!\! dt \, e^{-\imath \omega t} \, f(t) \, ,
\end{equation}
and where $\tilde{\kappa}$ is the (quantum) FDR kernel.
Therefore, the problem is completely specified in terms of the damping kernel (or equivalently, the spectral-density function).

Given a \emph{stationary} damping kernel, the Langevin equation can then be expressed in the Laplace domain as
\begin{align}
\left[ s^2 \mathbf{M} + 2 s \hat{\boldsymbol{\gamma}}(s) + \mathbf{C} \right] \hat{\mathbf{X}}(s) &= [ s \mathbf{M} \mathbf{X}_0 + \mathbf{P}_0 ] + \hat{\boldsymbol{\xi}}(s)\, ,
\end{align}
where $\mathbf{P}=M\dot{\mathbf{X}}$ and $(\mathbf{X}_0, \mathbf{P}_0)$ correspond to the initial values at $t=0$.
Formally, the solutions can be easily found by inversion:
\begin{align}
\hat{\mathbf{X}}(s) &= \hat{\mathbf{G}}(s) \, [ s \mathbf{M} \mathbf{X}_0 + \mathbf{P}_0 ] + \hat{\mathbf{G}}(s) \, \hat{\boldsymbol{\xi}}(s)\, , \\
\hat{\mathbf{G}}(s) &= \left[ s^2 \mathbf{M} + 2 s \hat{\boldsymbol{\gamma}}(s) + \mathbf{C} \right]^{-1} \, . \label{eq:G(s)}
\end{align}
Note that since our damping kernel is symmetric, i.e. $\boldsymbol{\gamma}(t,\tau) = \boldsymbol{\gamma}^\mathrm{T}(t,\tau)$, the same will be true for the propagator $\mathbf{G}(t,\tau)$ and its Laplace transform.
It is also useful to consider the following representation:
\begin{align}
\hat{\mathbf{G}}(s) &= \mathbf{M}^{-\frac{1}{2}} \left[ s^2 + 2 s \, \hat{\boldsymbol{\lambda}}(s) + \boldsymbol{\Omega}^2 \right]^{-1} \mathbf{M}^{-\frac{1}{2}} \, , \label{eq:G(s)M} \\
\hat{\boldsymbol{\lambda}}(s) & \equiv \mathbf{M}^{-\frac{1}{2}} \hat{\boldsymbol{\gamma}}(s) \, \mathbf{M}^{-\frac{1}{2}} \, , \\
\boldsymbol{\Omega}^2 & \equiv \mathbf{M}^{-\frac{1}{2}} \mathbf{C} \, \mathbf{M}^{-\frac{1}{2}} \, ,
\end{align}
where the eigenvalues of $\boldsymbol{\Omega}^2$ coincide with the squared frequencies of the normal modes of the free system.
Back in the time domain we have
\begin{align}
\mathbf{X}(t) &= \dot{\mathbf{G}}(t) \, \mathbf{M} \, \mathbf{X}_0 + \mathbf{G}(t) \,\mathbf{P}_0 + (\mathbf{G} * \boldsymbol{\xi})(t)\, ,
\end{align}
with $*$ denoting the Laplace convolution, defined as
\begin{equation}
(A*B)(t) = \int_0^t \!\! d\tau \, A(t\!-\!\tau) \, B(\tau).
\end{equation}

At this point it is worth pointing out that prior to \cite{QBM,Anastopoulos10},
non-perturbative $N$-oscillator master equations have been limited to the parameter regimes in which the $N$-oscillator problem was equivalent to independent single-oscillator problems.
More specifically, either the coupling was chosen such that $\hat{\boldsymbol{\gamma}} \propto \mathbf{1}$ \cite{Chou08} or the 
the resonance regime with $\mathbf{C} \propto \mathbf{1}$ was considered \cite{Chou08}, or both \cite{Paz08}.
It is more or less obvious from the Langevin equation that these examples, where $\hat{\boldsymbol{\gamma}}$ and $\mathbf{C}$ commute, are all trivial.

\subsection{Phase Space: The Solutions}
The compact and powerful formalism developed in Ref.~\cite{QBM} and which allowed a quick derivation of the master equation and its general solution, also makes straightforward the generalization to $N$ system oscillators. First, one introduces the following phase-space representation
\begin{eqnarray}
\mathbf{Z}^\mathrm{T} &\equiv& \left( \mathbf{X} , \mathbf{P} \right) \, , \\
\boldsymbol{\Xi}^\mathrm{T} &\equiv& \left( \mathbf{0} , \boldsymbol{\xi} \right) \, , \\
\mathbf{N}(t,\tau) &\equiv& \left[ \begin{array}{cc} \mathbf{0} & \mathbf{0} \\ \mathbf{0} & \boldsymbol{\nu}(t,\tau) \end{array} \right] \, , \\
\boldsymbol{\Gamma}(t,\tau) &\equiv& \left[ \begin{array}{cc} \mathbf{0} & \mathbf{0} \\ \mathbf{0} & \boldsymbol{\gamma}(t,\tau) \end{array} \right] \, .
\end{eqnarray}
The solutions of the Langevin equation can then be expressed as
\begin{eqnarray}
\mathbf{Z}(t) &=& \boldsymbol{\Phi}(t) \mathbf{Z}_0 + (\boldsymbol{\Phi} * \boldsymbol{\Xi})(t)\, , \label{eq:solution} \\
\boldsymbol{\Phi}(t) &=& \left[ \begin{array}{cc} \dot{\mathbf{G}}(t) \mathbf{M} & \mathbf{G}(t) \\ \mathbf{M} \ddot{\mathbf{G}}(t) \mathbf{M} & \mathbf{M} \dot{\mathbf{G}}(t) \end{array} \right] \, . \label{eq:matrix_prop}
\end{eqnarray}
As shown in Ref.~\cite{CRV03}, the reduced Wigner function of the system can be conveniently written as a double average over the stochastic source and the initial conditions:
\begin{equation}
W(\mathbf{Z},t)
= \Big\langle \Big\langle \delta \big( \mathbf{Z}(t) - \mathbf{Z} \big) \,
\Big\rangle_{\!\boldsymbol{\Xi}} \Big\rangle_{\mathbf{Z}_0}
\label{eq:average0} ,
\end{equation}
where $\mathbf{Z}(t)$ is  a solution of the Langevin equation as given by \eqref{eq:solution}.
It is then convenient to consider the Fourier transform from phase-space coordinates $\mathbf{Z}$ to the conjugate Fourier-domain coordinates $\mathbf{K}$
\begin{align}
\mathcal{W}(t,\mathbf{K}) &= \iint \! d^{N}\!X \, d^{N}\!P \, e^{-\imath \, \mathbf{K}^\mathrm{T} \mathbf{Z}} \, W(t,\mathbf{Z}) \, ,
\end{align}
which corresponds to the characteristic function from which the Weyl-ordered correlation functions can be directly obtained by functionally differentiating with respect to $\mathbf{K}$. Following the same procedure as in Sec.~2.3 of Ref.~\cite{QBM}, one can immediately get
\begin{align}
\mathcal{W}(t,\mathbf{K}) &= \mathcal{W}(0,\boldsymbol{\Phi}^\mathrm{T}\!(t)\mathbf{K}) \, e^{-\frac{1}{2} \mathbf{K}^\mathrm{T} \boldsymbol{\sigma}_T(t) \mathbf{K} } \, ,
\end{align}
where $\boldsymbol{\sigma}_T(t)$ is the thermal covariance matrix
\begin{align}
\boldsymbol{\sigma}_T(t) &= \int_0^t \!\! d\tau \int_0^t \!\! d\tau' \, \boldsymbol{\Phi}(t\!-\!\tau) \mathbf{N}(\tau,\tau') \boldsymbol{\Phi}^\mathrm{T}\!(t\!-\!\tau') \, . \label{eq:TS}
\end{align}
Since the Wigner function is a pseudo-distribution (it is real, normalized, but not necessarily everywhere-positive) its Fourier transform is a characteristic (or moment generating) function.
A straightforward interpretation of the solution immediately follows: the initial cumulants of the system experience damped oscillations due to the homogeneous propagator $\boldsymbol{\Phi}(t)$ while a thermal covariance arises,
smearing out details of the dissipating initial state.

The thermal covariance can also be broken down into the block-matrix correlations
\begin{align}
\boldsymbol{\sigma}_T^{\mathbf{XX}}\!(t) &= \int_0^t \!\! d\tau \! \int_0^t \!\! d\tau' \, \mathbf{G}(t\!-\!\tau) \boldsymbol{\nu}(\tau,\tau') \mathbf{G}^\mathrm{T}(t\!-\!\tau') \, , \label{eq:TSxx} \\
\boldsymbol{\sigma}_T^{\mathbf{PX}}\!(t) &= \int_0^t \!\! d\tau \! \int_0^t \!\! d\tau' \, \mathbf{M} \dot{\mathbf{G}}(t\!-\!\tau) \boldsymbol{\nu}(\tau,\tau') \mathbf{G}^\mathrm{T}(t\!-\!\tau') \, , \\
\boldsymbol{\sigma}_T^{\mathbf{XP}}\!(t) &= \left[ \boldsymbol{\sigma}_T^{\mathbf{PX}}(t) \right]^\mathrm{T} \, , \\
\boldsymbol{\sigma}_T^{\mathbf{PP}}\!(t) &= \int_0^t \!\! d\tau \! \int_0^t \!\! d\tau' \, \mathbf{M} \dot{\mathbf{G}}(t\!-\!\tau) \boldsymbol{\nu}(\tau,\tau') \dot{\mathbf{G}}^\mathrm{T}(t\!-\!\tau') \mathbf{M} \, . \label{eq:TSpp}
\end{align}
As explained above, we have $\mathbf{G}^\mathrm{T}(t,\tau) = \mathbf{G}(t,\tau)$, so that we do not need to consider the transpose of the propagator in Eqs.~\eqref{eq:TSxx}-\eqref{eq:TSpp}.

Strictly speaking, since we already have the solutions, we do not need any master equation, but we provide it for completeness. The derivation of Ref.~\cite{QBM} can be staightforwardly generalized to the case of $N$ system oscillators and the resulting master equation can be written as
\begin{align}
\frac{\partial}{\partial t} W(\mathbf{Z};t) &= \left\{ \boldsymbol{\nabla}_{\mathbf{Z}}^{\mathrm{T}} \, \boldsymbol{\mathcal{H}}(t) \, \mathbf{Z} + \boldsymbol{\nabla}_{\mathbf{Z}}^{\mathrm{T}} \, \mathbf{D}(t) \, \boldsymbol{\nabla}_{\mathbf{Z}} \right\} W(\mathbf{Z};t) , 
\end{align}
given the coefficient matrices
\begin{align}
\boldsymbol{\mathcal{H}}(t) &\equiv -\dot{\boldsymbol{\Phi}}(t) \, \boldsymbol{\Phi}^{-1}(t) , \\
\mathbf{D}(t) &\equiv \frac{1}{2} \left\{ \boldsymbol{\mathcal{H}}(t) \, \boldsymbol{\sigma}_T(t) + \boldsymbol{\sigma}_T(t) \, \boldsymbol{\mathcal{H}}^{\!\mathrm{T}}\!(t) + \dot{\boldsymbol{\sigma}}_T(t) \right\} , \label{eq:Ddynamic}
\end{align}
where $\boldsymbol{\mathcal{H}}$ are the homogeneous coefficients which contain system renormalization and dissipation,
and $\mathbf{D}$ are the diffusive and anti-diffusive coefficients.
(Note that the invalidity of previous derivations of the master equation for nonlocal dissipation pointed out in Ref.~\cite{QBM} applies in the same way to this case.)

\section{Refinement of the Solutions}\label{sec:refine}
\subsection{Local Damping}
For local dissipation it is very easy to solve for the phase-space propagator,
as it is formally determined by the matrix exponential
\begin{eqnarray}
\boldsymbol{\Phi}(t) &=& e^{-t\,\boldsymbol{\mathcal{H}}} \, , \\
\boldsymbol{\mathcal{H}} &=& \left[ \begin{array}{cc} \mathbf{0} & -\mathbf{M}^{-1} \\ \mathbf{C} & 2 \boldsymbol{\gamma}_0 \mathbf{M}^{-1} \end{array} \right] \, .
\end{eqnarray}
Directly solving for the position propagator involves an equivalent but quadratic matrix equation.
Even for large systems, diagonalization of $\boldsymbol{\mathcal{H}}$ perturbatively in $\boldsymbol{\gamma}_0$ is straightforward.

\subsection{Rational Damping}
\subsubsection{Pseudo-normal Modes}
If one has a rational damping kernel in the Laplace domain (or even meromorphic), then the propagator will also be rational (or meromorphic) in the Laplace domain.
Therefore, it can be decomposed into simple fractions which correspond to \emph{pseudo-normal} modes: 
\begin{align}
\hat{\mathbf{G}}(s) &= \mathbf{M}^{-\frac{1}{2}} \left[ \sum_k \frac{\frac{1}{f_k}}{s-f_k} \mathbf{U}_{\!k} \, \mathbf{U}_{\!k}^\dagger \right] \mathbf{M}^{-\frac{1}{2}} \, , \\
\mathbf{G}(t) &= \mathbf{M}^{-\frac{1}{2}} \left[ \sum_k \frac{e^{f_k t}}{f_k} \, \mathbf{U}_{\!k} \, \mathbf{U}_{\!k}^\dagger \right] \mathbf{M}^{-\frac{1}{2}} \, ,
\end{align}
assuming no repeated roots,
and where the $\mathbf{U}_{\!k}$ form an overcomplete basis given the initial condition $\dot{\mathbf{G}}(0) = \mathbf{M}^{-1}$, symmetry and the uniqueness of solutions.
These are not true modes and the $\mathbf{U}_{\!k}$ are not orthonormal because for nonlocal damping the pseudo-modes outnumber the system degrees of freedom.
However, they will be associated with non-physical modes in a higher-dimensional linear system in the following subsection.

From the frequency representation of Eq.~\eqref{eq:G(s)M} and the Langevin equation, the pseudo-mode decomposition must satisfy
\begin{align}
\left[ s^2 + 2 s \, \hat{\boldsymbol{\lambda}}(s) + \boldsymbol{\Omega}^2 \right] \left[ \sum_k \frac{\frac{1}{f_k}}{s-f_k} \mathbf{U}_{\!k} \, \mathbf{U}_{\!k}^\dagger \right]  &= \mathbf{1} \, ,
\end{align}
for all $s$.
Taking the limit $s \to f_k$ reveals a necessary condition for convergence, the characteristic equation:
\begin{align}
\left[ f^2 + 2 f \, \hat{\boldsymbol{\lambda}}(f) + \boldsymbol{\Omega}^2 \right] \mathbf{U}  &= \mathbf{0} \, , \label{eq:NonlocalChar}
\end{align}
which is a nonlinear eigen-value equation.
Canonical-like perturbation theory applied to this equation will be in agreement with results from the second-order master equation \cite{QOS},
but this method is much more efficient in calculating these particular frequencies to higher order as this method does not require any integration.

\subsubsection{Rational Damping: Extended Phase Space}
For the nonlocal damping kernel which is rational in Laplace space, we can transform the nonlocal problem into an effectively time-local and time-homogeneous one by considering a higher dimensional phase space.
This was possible for one system oscillator, but is much more useful with $N$ oscillators.
The specific problem that we will work out will be the regulated Ohmic coupling
\begin{eqnarray}
\hat{\boldsymbol{\gamma}}(s) &=& \frac{\boldsymbol{\gamma}_0}{ 1 + \frac{s}{\boldsymbol{\Lambda}} } \, ,
\end{eqnarray}
where the Ohmic-limiting damping matrix $\boldsymbol{\gamma}_0$ and the cutoff matrix $\boldsymbol{\Lambda}$ commute and are both positive definite.
This damping kernel is a rational function of order $[0/1]$ in $s$ and corresponds to 
\begin{eqnarray}
\tilde{\boldsymbol{\gamma}}(\omega) &=&  \frac{2 \, \boldsymbol{\gamma}_0}{ 1 + \left(\frac{\omega}{\boldsymbol{\Lambda}}\right)^2 } \, ,
\end{eqnarray}
in the Fourier domain; see Eq.~\eqref{eq:damping2} below.
For this to be a valid coupling, one only needs to ensure that $\tilde{\boldsymbol{\gamma}}(\omega) > \mathbf{0}$ for all $\omega$, which is indeed the case.

Together with the initial conditions $(\mathbf{X}_0,\mathbf{P}_0)$ Eq.~\eqref{eq:langevin2} yields the following \emph{initial-value constraints}:
\begin{eqnarray}
\mathbf{X}(0) &=& \mathbf{X}_0 \, , \\
\mathbf{M}\dot{\mathbf{X}}(0) &=& \mathbf{P}_0 \, , \\
\mathbf{M}\ddot{\mathbf{X}}(0) &=& -\underbrace{\left[\mathbf{C} + 2 \boldsymbol{\Lambda} \boldsymbol{\gamma}_0 \right]}_{\mathbf{C}_\mathrm{\!bare}} \mathbf{X}_0 \, , \label{eq:F(0)}
\end{eqnarray}
where the initial acceleration will always appear to have a bare or non-renormalized frequency.
In general, if the damping kernel is of order $[n/d]$ then we will need to determine $d$ additional initial conditions beyond the first two.

Next, we factor out the damping kernel's denominator and represent the homogeneous Langevin equation in Laplace space as a polynomial
\begin{align}
& \left[ \left( 1 + s \, \boldsymbol{\Lambda}^{\!-1} \right) \left( s^2 \mathbf{M} + \mathbf{C} \right) + 2 s \, \boldsymbol{\gamma}_0 \right] \hat{\mathbf{X}}(s) \nonumber \\
& = \left( 1 + s \, \boldsymbol{\Lambda}^{\!-1} \right) \left( s\, \mathbf{M} \mathbf{X}_0 + \mathbf{P}_0 \right) \, ,
\end{align}
which is equivalent to the third-order differential equation 
\begin{align}
& \left\{ \boldsymbol{\Lambda}^{\!-1} \mathbf{M} \frac{d^3}{dt^3} + \mathbf{M} \frac{d^2}{dt^2} + \left( 2 \boldsymbol{\gamma}_0 \!+\! \boldsymbol{\Lambda}^{\!-1} \mathbf{C} \right)\! \frac{d}{dt} + \mathbf{C} \right\} \mathbf{X}(t)  \nonumber \\
& = \mathbf{0} \, ,
\end{align}
with constant coefficients.
Given that we have made no approximations, solutions to this equation with proper initial conditions are immune to any issues of runaway solutions or causality violation often associated with higher-order differential equations.

Because we have a linear system of ordinary differential equations, we can solve this system with a matrix exponential analogous to the case of local damping.
Here it is natural to consider the extension of phase space into ``forces'' such that
\begin{equation}
\frac{d}{dt} \left[ \begin{array}{c} \mathbf{X} \\ \mathbf{P} \\ \mathbf{F} \end{array} \right] =
\left[ \begin{array}{ccc} \mathbf{0} & \mathbf{M}^{-1} & \mathbf{0} \\ \mathbf{0} & \mathbf{0} & \mathbf{1} \\ -\boldsymbol{\Lambda} \mathbf{C} & (-2 \boldsymbol{\Lambda} \boldsymbol{\gamma}_0 \!-\! \mathbf{C}) \mathbf{M}^{-1} & - \boldsymbol{\Lambda} \end{array} \right]
\left[ \begin{array}{c} \mathbf{X} \\ \mathbf{P} \\ \mathbf{F} \end{array} \right] \, , \label{eq:ForcePhaseSpace}
\end{equation}
and let us denote this effective time-translation generator as $\boldsymbol{\mathcal{F}}$ so that we can write our solutions as
\begin{eqnarray}
\left[ \begin{array}{c} \mathbf{X}(t) \\ \mathbf{P}(t) \\ \mathbf{F}(t) \end{array} \right] &=& e^{t \, \boldsymbol{\mathcal{F}}} \left[ \begin{array}{c} \mathbf{X}_0 \\ \mathbf{P}_0 \\ \mathbf{F}_0 \end{array} \right] \, ,
\end{eqnarray}
where the initial condition $\mathbf{F}_0$ is determined by Eq.~\eqref{eq:F(0)}.
Given our initial value constraints, we can then express our solutions as
\begin{equation}
\left[ \begin{array}{c} \mathbf{X}(t) \\ \mathbf{P}(t) \\ \mathbf{F}(t) \end{array} \right] = e^{t \, \boldsymbol{\mathcal{F}}} \left[ \begin{array}{ccc} \mathbf{1} & \mathbf{0} & \mathbf{0} \\ \mathbf{0} & \mathbf{1} & \mathbf{0} \\ - 2 \boldsymbol{\Lambda} \boldsymbol{\gamma}_0 \!-\! \mathbf{C} & \mathbf{0} & \mathbf{0} \end{array} \right] \left[ \begin{array}{c} \mathbf{X}_0 \\ \mathbf{P}_0 \\ \mathbf{0} \end{array} \right] \, . \label{eq:IVC}
\end{equation}
The initial-value constraint matrix, which will be denoted by $\mathbf{T}$,
maps the two initial conditions $\left( \mathbf{X}_0 , \mathbf{P}_0 , \mathbf{0} , \cdots \right)$ of the nonlocal equation into the larger number of initial conditions required by the effective local equation $\left( \mathbf{X}_0 , \mathbf{P}_0 , \mathbf{F}_0 , \cdots \right)$.%
\footnote{In general, the dimension of the ``extended phase space'' and the rank of $\mathbf{T}$ will correspond to the order of the equivalent ordinary differential equation.}
Given this representation we can now identify the $2 \times 2$ leading principal minor of $e^{t \, \boldsymbol{\mathcal{F}}} \, \mathbf{T}$ with $\boldsymbol{\Phi}(t)$, since this matrix would map $\left( \mathbf{X}_0 , \mathbf{P}_0 \right)$ onto $\big( \mathbf{X}(t) , \mathbf{P}(t) \big)$.
In the particular example that we are considering here $\mathbf{F}_0$ has no  dependence on $\mathbf{P}_0$ and the second column of $e^{t \, \boldsymbol{\mathcal{F}}}$ directly yields $\left( \mathbf{G}(t), \mathbf{M} \dot{\mathbf{G}}(t), \mathbf{M} \ddot{\mathbf{G}}(t) \right)$, so that one can read the result for the propagator and its first two derivatives and obtain $\boldsymbol{\Phi}(t)$ making use of Eq.~\eqref{eq:matrix_prop}. Hence, in the end one has been able to find the matrix propagator for the integro-differential equation by solving a higher-order linear differential equation.

\subsection{Late-Time Thermal Covariance}
Using the results in Appendix~E of Ref.~\cite{QBM}, we can reduce the late-time covariance to a single frequency integral in the Laplace domain:
\begin{align}
\hat{\boldsymbol{\sigma}}_T(s) & \approx \frac{1}{s} \frac{1}{2\pi} \int_{\!-\infty}^{+\infty} \!\!\! d\omega \, \hat{\boldsymbol{\Phi}}(s\!+\!\imath \omega) \,\tilde{\mathbf{N}}(\omega) \, \hat{\boldsymbol{\Phi}}^{\!\mathrm{T}}\!(s\!-\!\imath \omega) \, , 
\end{align}
which is exact in the late-time limit
\begin{equation}
\boldsymbol{\sigma}_T(\infty) = \lim_{s \to 0} s \, \hat{\boldsymbol{\sigma}}_T(s) \, ,
\end{equation}
and for all times if the dissipation is local.

Moreover, there is a simplification to this expression which was not exceptionally useful for a single system mode but is extremely useful for multiple system modes.
The expression can be further simplified from a quadratic form into a linear form.
First we express the noise kernel in terms of the damping kernel by means of the FDR \eqref{eq:FDR}.
Next we relate the Fourier transform of the damping kernel to its Laplace transform by first noting
\begin{align}
\tilde{\boldsymbol{\gamma}}(\omega) &= \int_{-\infty}^{+\infty} \!\!\! dt \, e^{-\imath \omega t} \, \boldsymbol{\gamma}(t) \, , \\
&= \int_0^{\infty} \!\! dt \, e^{-\imath \omega t} \, \boldsymbol{\gamma}(t) + \int_0^{\infty} \!\! dt \, e^{+\imath \omega t} \, \boldsymbol{\gamma}(-t) \, .
\end{align}
Then we use the fact that the damping kernel is double-Hermitian (see Ref.~\cite{QOS}) to obtain
\begin{align}
\tilde{\boldsymbol{\gamma}}(\omega) &= \int_0^{\infty} \!\! dt \, e^{-\imath \omega t} \, \boldsymbol{\gamma}(t) + \int_0^{\infty} \!\! dt \, e^{+\imath \omega t} \, \boldsymbol{\gamma}^\mathrm{\!T}\!(t) \, .
\end{align}
Assuming no singularities in $\hat{\boldsymbol{\gamma}}(s)$ for Re$[s]>0$, which should be the case as the damping kernel is a somewhat localized distribution in time,
we may analytically continue the above relation to obtain
\begin{eqnarray}
\tilde{\boldsymbol{\gamma}}(\omega) &=& \lim_{s \to 0} \left[ \hat{\boldsymbol{\gamma}}(s \!+\! \imath \omega) + \hat{\boldsymbol{\gamma}}^\mathrm{\!T}\!(s \!-\! \imath \omega) \right] \, .
\label{eq:damping2}
\end{eqnarray}
Then we expand $\mathrm{Re}[\hat{\boldsymbol{\gamma}}(\imath \omega)]$ pairing $\hat{\boldsymbol{\gamma}}(+\imath \omega)$ with $\hat{\boldsymbol{\Phi}}(+\imath \omega)$ and $\hat{\boldsymbol{\gamma}}^\mathrm{\!T}\!(-\imath \omega)$ with $\hat{\boldsymbol{\Phi}}^{\!\mathrm{T}}\!(-\imath \omega)$.
One can then apply the Langevin equation evaluated at imaginary frequencies to eliminate any explicit dependence on the damping kernel, so that only the implicit dependence of the propagator is left.
After some simplification and application of symmetries
we are left with the expression
\begin{align}
\boldsymbol{\sigma}_T^\infty &= \frac{1}{2\pi} \int_{\!-\infty}^{+\infty} \!\!\! d\omega \, \frac{\tilde{\kappa}(\omega)}{\omega} \, \mathrm{Im}\!\left[ \begin{array}{cc} \hat{\mathbf{G}}(-\imath \omega) & \mathbf{0} \\ \mathbf{0} & \omega^2 \, \mathbf{M} \, \hat{\mathbf{G}}(-\imath \omega) \, \mathbf{M} \end{array} \right] , \label{eq:sigma(inf)} 
\end{align}
which avoids all phase-space matrix multiplication.

\section{System Modes in a Common Environment}\label{sec:Detectors}
Here we consider identical Unruh-DeWitt detectors $Q_i$ in a shared environment (a massless scalar field in a vacuum or thermal state), but at different locations $\mathbf{r}_i$ as considered in Ref.~\cite{Lin09}.
The coupling is local and linear between the system positions (more typically momenta) and field operator.
In this model the damping kernel is given by
\begin{eqnarray}
\tilde{\gamma}_{ij}(\omega) &=& 2 \gamma_0 \, \mathrm{sinc}(r_{\!ij} \omega) \, , \\
\hat{\gamma}_{ij}(s) &=& \gamma_0 \, \frac{1-e^{-r_{ij} s}}{r_{\!ij} s} \, , \label{eq:gamma(s)}
\end{eqnarray}
in the Fourier and Laplace domains respectively,
and where $\mathbf{r}_{\!ij} = \mathbf{r}_i \!-\! \mathbf{r}_j$.
The Markovian limit is reached here when the environment is at high temperature and the detectors are taken to be very far apart,
(in which case one recovers $\gamma_0$ as the damping rate of the individual system oscillators)
or very close together (in which case one has effectively local cross-damping).
At finite temperature it is well known that QBM requires a finite cutoff regulator for Ohmic coupling to the environment,
and this is also the case here.
Note that distance variation naturally takes the form of an Ohmic regulator in the limit of small separations.
Following Ref~\cite{Dipole}, which considered the same environment,
we will not evaluate the correlations more precisely than to some smallest scale $r_0$ in the sense that we take
\begin{equation}
\lim_{\mathbf{r}_i \to \mathbf{r}_j} r_{\!ij} \,=\, r_0 \,\equiv\, \Lambda^{-1} \, ,
\end{equation}
to provide a natural cutoff of $\Lambda$. The scale $r_0$ can be interpreted as the effective size of the detectors. A more detailed analysis would require including the form factors for the detectors, but the conclusions should be qualitatively the same.

For a pair of identical detectors $Q_1$ and $Q_2$,
the dynamics is naturally described in terms of the sum and difference $Q_\pm = Q_1 \pm Q_2$ (which correspond to the normal modes of the free theory) and a similar combination for the components of the damping kernels,
\begin{eqnarray}
\hat{\gamma}_\pm(s) &=& \hat{\gamma}_{11}(s) \pm \hat{\gamma}_{12}(s) \, ,
\end{eqnarray}
as well as the noise kernel (by the FDR).
Hence, when the two detectors are very close, their difference $Q_-$ experiences vanishing dissipation.
It is associated with a decoherence-free subspace, involving \emph{dark states},
but only in the limit of minimal separation so that their environments appear identical and are perfectly correlated.
This is essentially the regime considered in Ref.~\cite{Paz08} for the entanglement dynamics of two oscillators in a common environment.
But more generally, for small separations the sum $Q_+$ relaxes quickly and then on a much longer timescale the difference $Q_-$ will also thermalize.
The resultant asymptotic state, which is Gaussian in the sum and difference, can also be entangled at higher order in the coupling.

More generally Ref.~\cite{Lin09} considered finite separation distances and determined the oscillators to be asymptotically entangled when near the cutoff scale in separation.
Our more general formalism allows for non-resonant detectors which cannot be transformed into a pair of individual quantum Brownian oscillators.
Hence, it enables us to determine how resonant the detectors must be for this asymptotic entanglement to ensue.

\subsection{Regulation and Integration}
\begin{figure}[h]
\centering
\includegraphics[width=0.5\textwidth]{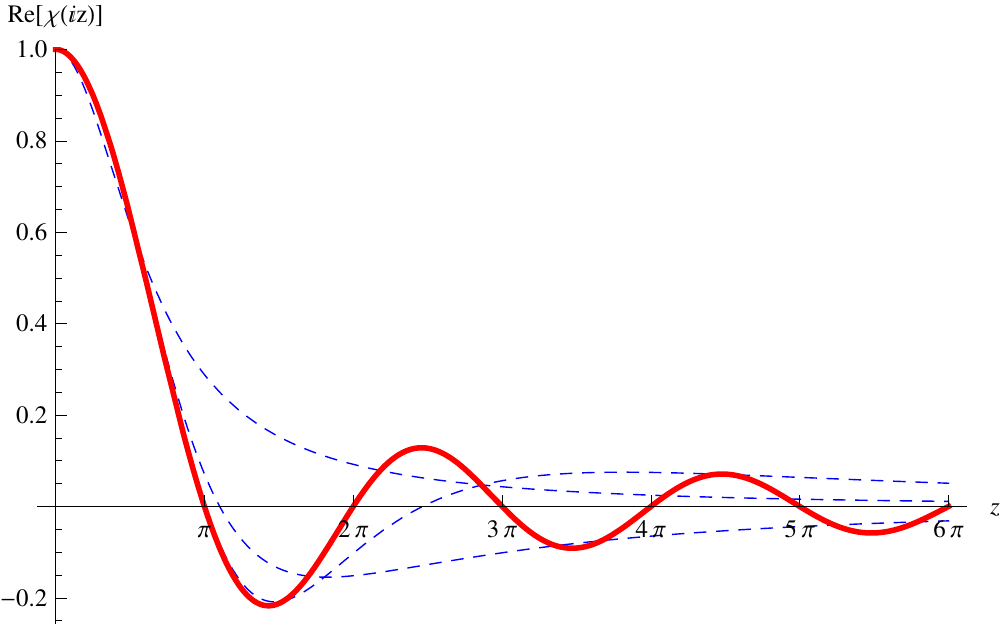}
\caption{\textcolor{blue}{$\cdots$ The first three $\chi_{[n/n+1]}(z)$ Pad\'{e} approximants} of the \textcolor{red}{$\bullet$ exact regulator}.}
\label{fig:Pade}
\end{figure}

The non-perturbative late-time covariance \eqref{eq:sigma(inf)} evaluated with our exponential cutoff-regulator \eqref{eq:gamma(s)}, is exceptionally oscillatory
and does not lend itself to well-behaved numerics, especially in the near regime.
In evaluating the late-time covariance, the exact regulator in the Laplace domain
\begin{align}
\chi(z) &= \frac{1-e^{-z}}{z} \, ,
\end{align}
must be evaluated at imaginary arguments, and it becomes both oscillatory and asymptotically $\mathcal{O}(1/\imath z)$ as $|z| \to \infty$.
Therefore, the appropriate small-$r$ approximations of this regulator are the $\chi_{[n/n+1]}(z)$ Pad\'{e} approximants.
These are the best local (small-$z$ expansion) and rational (ratio of polynomials) approximations which are also $\mathcal{O}(1/\imath z)$. 
We plot the first three such approximants in Fig.~\ref{fig:Pade}.
One can see that they are good approximations up to $r \sim n \pi / 2 \Omega$, similar to the Taylor series convergence for sinusoidal functions.
In this work we have primarily considered the first-order Pad\'{e} approximant
\begin{align}
\chi_{[0/1]}(z) &= \frac{1}{1+\frac{z}{2}} \, ,
\end{align}
which is also an Ohmic regulator for small separations.
This local approximation is most accurate in the high-cutoff and small-$r$ regime, which is precisely where we need to investigate more carefully,
but it also has the correct asymptotics.
Resultant calculations can be obtained exactly and are very well behaved.
In addition to numeric solutions and analytic small-$r$ solutions, we also employ analytic small-$\gamma_0$ solutions using weak-coupling perturbation of Eq.~\eqref{eq:NonlocalChar}.

\subsection{Entanglement Dynamics}
In calculations of entanglement we consider both the Peres-Horodecki criterion $\Sigma$ \cite{Simon00} and the logarithmic negativity $E_{\mathcal{N}}$ \cite{Vidal02} as a consistency check.
In all graphs, positive values denote entanglement and negative ones separability.
We consider the zero-temperature regime only as it emphasizes entanglement.

Our generic analysis of entanglement dynamics involves two factors:
(1) the relevant timescales for decoherence and (2) \emph{unmaximized} entanglement monotones of the asymptotic state \cite{Cummings10}.
In this linear model decoherence arises mainly due to the growth of the thermal covariance $\boldsymbol{\sigma}_T$, which smears away oscillations in the Wigner function corresponding to quantum interference.
A more thorough explanation of the general evolution is given in Sec.~4 of Ref.~\cite{QBM}.

\begin{figure}[h]
\centering
\includegraphics[width=0.5\textwidth]{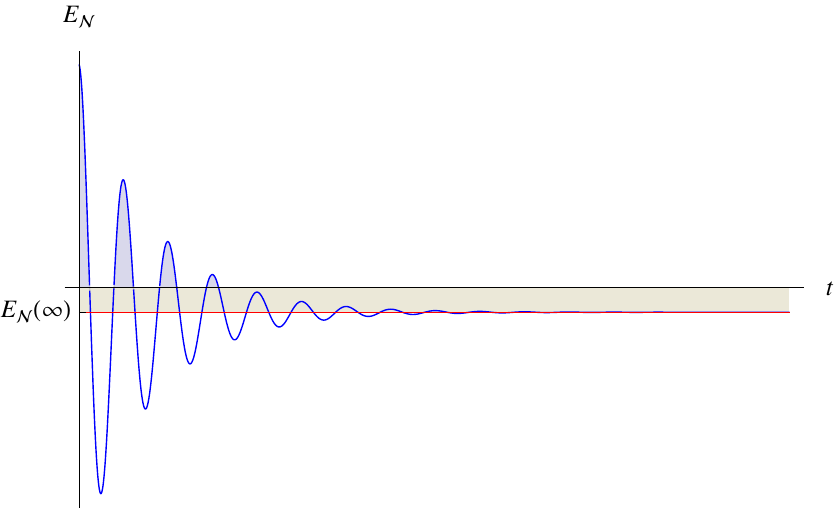}
\caption{\label{fig:EntDynamic}Qualitative plot of (unmaximized) entanglement dynamics including entanglement sudden death, revival, and asymptotic separability.}
\end{figure}
Traditionally defined entanglement monotones are not sufficient to give a cursory analysis of entanglement sudden death as they do not distinguish between separable states.
Given some operationally defined entanglement monotone $\mathcal{E} = \max[ 0, \underline{\mathcal{E}} ]$, so that $\mathcal{E}$ is zero for all separable states,
then the unmaximized function $\underline{\mathcal{E}}$ is more useful when its evolution is continuous.
As we qualitatively plot in Fig.~\ref{fig:EntDynamic}, entanglement sudden death occurs because the unmaximized entanglement monotone asymptotes towards a negative value,
whereas the traditionally defined entanglement monotone does not asymptote towards zero.
Therefore, given that decoherence in linear models is fairly well understood,
we will focus primarily upon analysis of the asymptotic (unmaximized) entanglement monotone.

\subsubsection{Close Detectors}
\begin{figure}[h]
\centering
\includegraphics[width=0.5\textwidth]{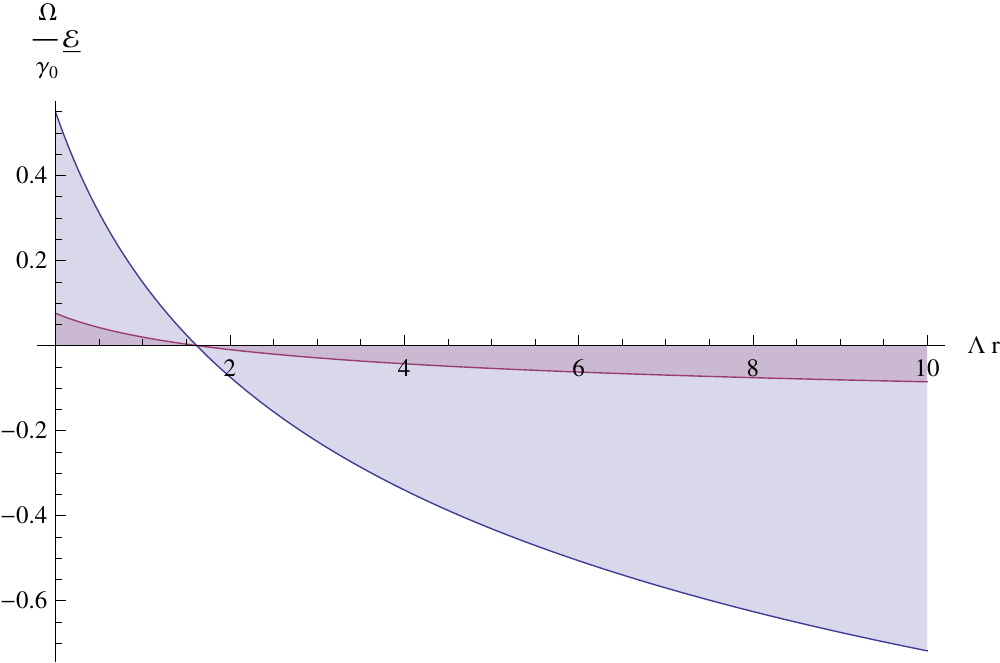}
\caption{\label{fig:EntR}The asymptotic (unmaximized) entanglement monotones of two resonant oscillators as a function of separation where $\gamma_0 = \Omega/10$ and $\Lambda = 100 \, \Omega$.}
\end{figure}
In agreement with Ref.~\cite{Lin09} we find the resonant oscillators to be asymptotically entangled when their separation is comparable to the inverse of the cutoff scale.
Fig.~\ref{fig:EntR} denotes the asymptotic entanglement measures of the resonant oscillators on the near cutoff scale where asymptotic entanglement can emerge.
It is curious that, despite the same environmental correlations and despite the emergence of a dark state under the same conditions,
for a pair of two-level atoms the possibility of asymptotic entanglement does not emerge (at second-order) for perfect cross-correlations \cite{Dipole}.
Moreover the behavior is rather opposite to that here, with the unmaximized entanglement monotones becoming more negative at proximity.

\subsubsection{Off-Resonant Detectors}
We know that asymptotic entanglement will ensue when the two oscillators are very close together and at resonance.
Therefore the question arises, how close and how at resonance must the two oscillators be to have asymptotic entanglement.
The first question has been answered: to have asymptotic entanglement, the oscillators must be extremely close together, of the order of the inverse cutoff.
The second question requires a more general multivariate treatment, which we can provide with this formalism.

\begin{figure}[h]
\centering
\includegraphics[width=0.5\textwidth]{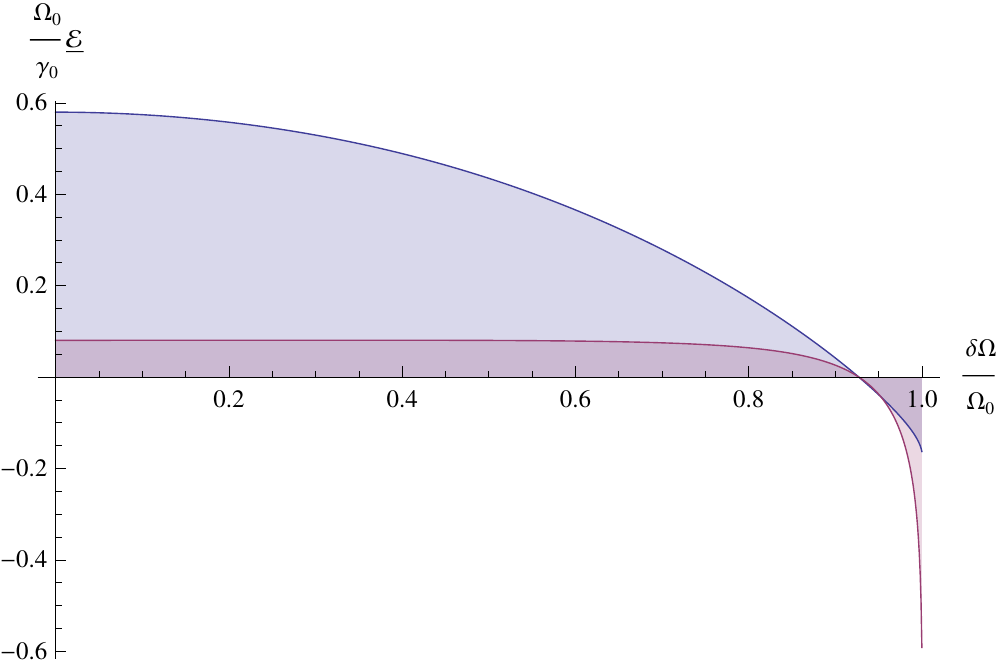}
\caption{\label{fig:EntD}The asymptotic entanglement measures of two oscillators as a function of detuning where $\gamma_0 = \Omega/10$ and $\Lambda = 100 \, \Omega$.}
\end{figure}

Here we take the two oscillators to be very close in position but with different frequencies $\Omega_\pm = \Omega_0 \pm \delta \Omega$.
Expressed in this manner with average frequency $\Omega_0$ and difference $2\delta \Omega$,
we find the asymptotic entanglement to be very much insensitive to $\delta \Omega$.
Fig.~\ref{fig:EntD} denotes the asymptotic entanglement measures of the two oscillators for a moderate coupling strength.
The stronger the environmental interaction is, the more asymptotic entanglement ensues and the less resonance sensitivity there is.
But even for weak interactions, resonance does not appear to be a very stringent criterion for asymptotic entanglement, at least far less stringent than proximity.

\subsection{Sub- and Super-radiance}
We know that the dark and bright states emerge when the two oscillators are very close together and at resonance.
Therefore, the following question arises: how close and how close to resonance must the two oscillators be to have sub- and super-radiance?

\begin{figure}[h]
\centering
\includegraphics[width=0.5\textwidth]{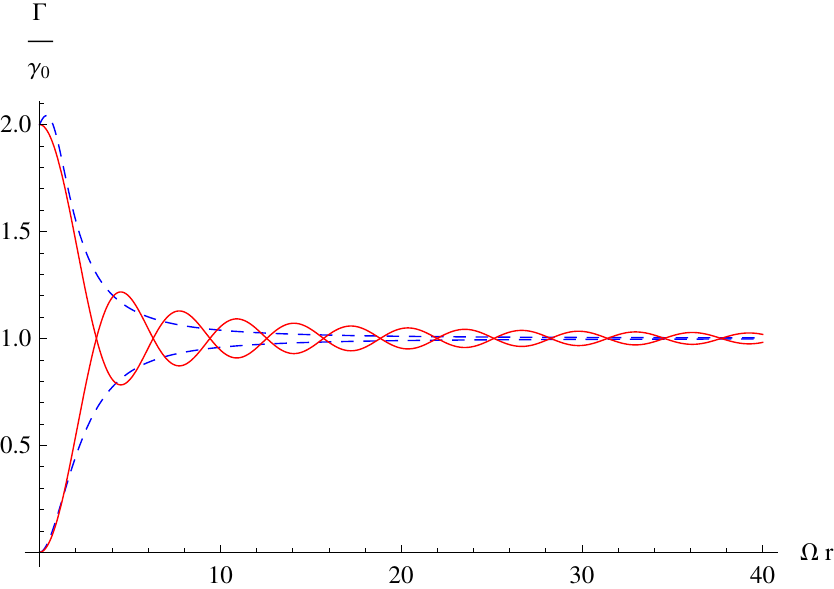}
\caption{\label{fig:SupSubR}Phenomenological decay rates as a function of separation for two resonant oscillators using \textcolor{red}{$\cdot$ weak-coupling perturbation} and \textcolor{blue}{$\cdots$ the small-$r$ Pad\'{e} approximation} where $\gamma_0 = \Omega/10$.}
\end{figure}

\begin{figure*}[h]
\centering
\includegraphics[width=\textwidth]{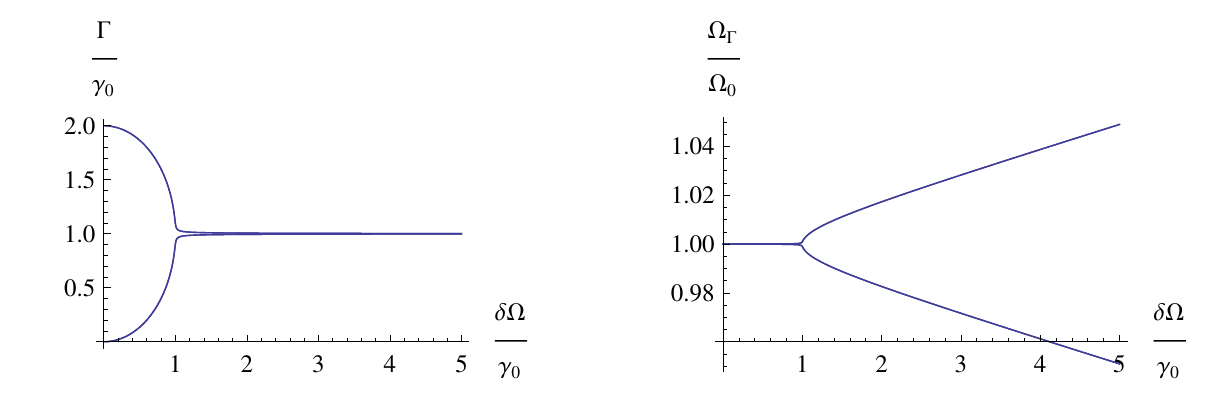}
\caption{\label{fig:SupSubD}Homogeneous timescales as a function of detuning for two close oscillators where $\gamma_0 = \Omega/100$.}
\end{figure*}

In Fig.~\ref{fig:SupSubR} we plot the phenomenological damping rates $\Gamma$ as a function of separation between two resonant oscillators.
The sub-radiant decay rate starts at zero while the super-radiant decay rate starts doubled.
Both then asymptote to the same value for large separations.
For the exact regulator, the intermediate behavior of the decay rates is somewhat oscillatory and the separation timescales play a fairly different (and more causal) role than a simple decay rate.
But the important information to extract is that the existence of the dark and bright states is not particularly sensitive to a lack of proximity (where our approximation is valid).
The relevant length scale appears to simply be the resonant wavelength.

In Fig.~\ref{fig:SupSubD} we plot the phenomenological damping and oscillation rates $\Gamma$ and $\Omega_\Gamma$ as a function of detuning between two close oscillators' free parameters.
To be more specific, these oscillators have effective modes which evolve with frequencies $\Gamma \pm \imath \, \tilde{\Omega}_\Gamma$ where $\tilde{\Omega}_\Gamma=\sqrt{\Omega_\Gamma^2-\Gamma^2}$.
As with two-level atoms \cite{Dipole}, the tuning needs to be better than the damping rate.
Here we can also see that the frequency of the damped oscillators initially resists any display of detuning until this critical threshold is reached.
Although the frequency $\Omega_\Gamma$ is very similar for the bright and dark modes, they oscillate at different rates since their damping rates (and hence the frequency $\tilde{\Omega}_\Gamma$) are very different.

\section{Discussion}
In this work we have considered the general model of a system of quantum oscillators bilinearly coupled to an environment of quantum oscillators.
We have shown how the phase-space solutions of Ref.~\cite{QBM} for a single oscillator can be easily extended to this more general case.
Moreover, we have provided useful relations for the multipartite case,
which greatly reduce the amount of necessary calculation.

We have applied our formalism to the problem of local oscillators interacting with a common scalar-field environment.
Our formalism is completely general and it has allowed us to explore this problem more thoroughly than before \cite{Paz08,Lin09}.
In particular we have found that detuning of the atoms does not have a profound effect upon the asymptotic entanglement.

In Ref.~\cite{QBM} we were able to provide exact, analytical solutions for various power-law spectral densities: sub-Ohmic, Ohmic, and supra-Ohmic.
Multivariate environment correlations arising from a quantum field are considerably more complex, because the spatial separation between the local oscillators causes a functional dependence for the generalized damping kernel which is far more complicated than an approximately power-law distribution.
In this work we have applied a three-way approach of (1) numerical integration of our simplified expressions, (2) weak-coupling perturbation
and (3) an appropriate Pad\'{e} approximation of the influence kernel.
Future work must push the boundaries of these approximation schemes and consider alternatives, such as the perturbative cross-correlation expansion
considered in Ref.~\cite{Lin09}.

\section*{Acknowledgments}
This work is supported in part by DARPA grant DARPAHR0011-09-1-0008 and the Laboratory for Physical Sciences.


\bibliography{bib}
\bibliographystyle{apsrev4-1}

\end{document}